\documentstyle[aps,preprint,psfig]{revtex}

\begin{document}

\title{Kinetic Slip Condition, van der Waals Forces, and Dynamic Contact Angle}
\author{
Len M. Pismen and  Boris Y. Rubinstein\\
\small \it Department of Chemical Engineering, Technion--Israel
Institute of Technology, Haifa 32000, Israel.}
\date{\today}
\maketitle

\begin{abstract}
The profiles of a spreading wetting film are computed taking into account intermolecular 
forces and introducing a kinetic slip condition at a molecular cut-off distance. This 
eliminates the stress singularity, so that both ``true" and ``visible" contact angles are 
defined unequivocally. The ``true" contact angle at the cut-off distance depends on the slip 
length as well as on the edge propagation speed, but not on gravity or asymptotic 
inclination angle. These macroscopic factors influence, however, the ``visible" contact 
angle observed in the interval where the actual film profile departs from the intermediate 
asymptotic curve.
\end{abstract}

\section{Introduction} \label{intro}

The two basic unsolved problems in the theory of a moving three-phase
contact line are defining the contact angle and resolving the infamous
viscous stress singularity. Different approaches to both problems, neither
of them satisfactory, have been reviewed by Shikhmurzaev \cite{shih}. Even
under equilibrium conditions, the structure of the three-phase region
cannot be understood without taking into account intermolecular
interactions between the fluid and the solid support \cite{Der,Isr,DGen85}.
It becomes apparent that motion of a contact line is an intrinsically {\it
mesoscopic\/} problem, and the dynamical theory should blend factors
contributed by hydrodynamics and physical kinetics.

The ``standard'' equilibrium contact angle $\theta_e$ is defined by the
Young--Laplace formula
\begin{equation}
\sigma_v - \sigma_l =\sigma \cos \theta_e.
\label{YL} \end{equation}
which involves surface tension $\sigma$ of an interface between two
semi-infinite fluid phases (in the simplest case, a one-component liquid
and its vapor) and (non-measurable) surface tensions between the solid
support and either fluid, $\sigma_l$ and $\sigma_v$. Since, by definition,
the standard surface tension refers to a boundary between semi-infinite
phases, the surface properties should be modified when the three-phase
region falls within the range of intermolecular forces, and therefore the
classical formula is likely to fail in a close vicinity of the contact
line. This region is too small to be detected by available measurement
techniques, but modification of interfacial properties is often revealed by
the formation of a precursor film. Thus, even under equilibrium conditions
the contact angle, generally, varies with the distance from the contact
line and cannot be defined unequivocally.

In a dynamical situation, such as wetting, spreading or draw-down of a
meniscus, the interfacial curvature, and hence, the change of the contact
angle, are further influenced by the viscous stress. A properly defined
contact angle fixes the boundary condition at the edge of an advancing or
receding film and is therefore a necessary ingredient for computation of
macroscopic flows, influenced also by external forces, such as gravity, and
by changes in temperature and chemical composition through buoyancy and
Marangoni effect. Macroscopic measurements yield the so-called ``visible''
contact angle, differing from both the ``standard'' value in Eq.~(\ref{YL})
and a hypothetical ``true'' (microscopic) value. Both ``true" and visible
contact angles should depend on the flow velocity and are subject to
hysteresis.

One should be warned that the very notion of a ``true'' interfacial angle
is precarious, since it extrapolates the concept of a sharp interface of a
continuous theory to molecular distances. This notion is eliminated
altogether in molecular simulations \cite{Kop89,Thom89} and in diffuse
interface theories \cite{amcf,keller,sepp,pp}. In continuum theories
incorporating intermolecular forces the ``true'' contact angle can be
defined {\it at most\/} at the molecular cut-off distance $d$. This is
sometimes forgotten when hydrodynamic theory leads to the appearance of
unphysically narrow boundary layers.

Bearing in mind limitations of continuum mechanics extended to molecular
scales, we attempt in this communication to combine the standard
hydrodynamic theory with a simple kinetic description of sliding motion in
the first molecular layer adjacent to the solid support. The thickness of
the sliding layer is identified with the cut-off length in the van der
Waals interaction potential; thus, the theory is expected to operate at
about the same crude level as the classroom derivation of the van der Waals
equation of state \cite{ll}.

The paper is organized as follows. We start in Section \ref{Slip} with a
detailed discussion of the slip condition. Basic equations in lubrication
approximation are formulated in Section \ref{Scale}. Intermediate
asymptotics of the solutions at relatively short macroscopic distances from
the contact  line, where gravity still does not come into play, are
discussed in Section \ref{Sasym}. Solutions describing the form of a
stationary meniscus on a moving inclined plane are given in Section
\ref{Smen}. 
%Our main qualitative result is demonstration of extreme
%sensitivity of the shape of a moving film at macroscopic distances from the
%contact line to molecular-scale factors operating in its immediate
%vicinity.

\section{Slip Condition} \label{Slip}

With any finite contact angle given as a boundary condition, a moving
contact line still cannot be described within the framework of conventional
hydrodynamics, since the classical no-slip condition on a solid substrate
generates a multivalued velocity and, hence, an infinite stress in the
vicinity of a contact line, leading formally to an infinite drag force
\cite{HuSc,DussDav74}.

The most common way to eliminate the viscous stress singularity is to
impose a phenomenological {\em slip} condition. Presence of slip at a
microscopic scale comparable with intermolecular distances is an
established fact in Maxwell's \cite{Maxwl} kinetic theory of gases; for
dense fluids it is a feasible hypothesis supported by molecular dynamics
simulations \cite{Kop89,Thom89}. The two alternatives are slip conditions
of ``hydrodynamic'' and ``kinetic'' type.

The version of the slip condition most commonly used in fluid-mechanical
theory is a linear relation between the velocity component along the solid
surface $u_{s}$ and the shear stress \cite{Lamb32,Bede76}. The
proportionality constant contains a phenomenological parameter -- slip
length
-- characterizing intermolecular interaction between the fluid and the
solid; in liquids this length should be small, so that the effect of
sliding becomes significant only in the vicinity of a moving contact line
where stresses are very large. This condition has been widely used for
modeling macroscopic flows involving the contact line motion
\cite{Hoc77,Grsp78,Hoc81,Hoc90,HalMiks91}. It does not eliminate the stress
singularity but only makes it integrable, thus leaving a logarithmic
(integrable) singularity of the interfacial curvature. This leads formally
to a breakdown of the commonly used lubrication approximation in the
vicinity of a contact line that can be remedied only by further {\it ad
hoc\/} assumptions, making the slip length dependent on the distance from
the contact line  \cite{Grsp78,HalMiks91}.

This drawback may be mere technical, but a more serious disadvantage of
hydrodynamic slip theories lies in their inherent inability to predict the
dynamic contact angle. Thus, the two basic problems become disentangled,
and, in addition to a phenomenological slip coefficient, empirical
relationships between the velocity and contact angle have to be introduced
in model computations.

Another version of the slip condition, rooted in physical kinetics
\cite{Blake69,Ruck79,DeG92,Ruck95}, defines the slip velocity through the
gradient of thermodynamic potential $w$ along the solid surface:
\begin{equation}
   u_{s}= - \frac{D}{n kT } \, \nabla w, \label{slide1}
\end{equation}
where $D$ is surface diffusivity, $n$ is particle number density, 
$k$ is Boltzmann
constant, $T$ is temperature, and $\nabla$ is two-dimensional gradient
operator along the solid surface. The condition (\ref{slide1}) follows
rather naturally from considering activated diffusion in the first
molecular layer adjacent to the solid. In contrast to the hydrodynamic slip
condition, the kinetic condition (\ref{slide1}) can be used to define the
``true'' dynamic contact angle at the contact line in a unique  way, as we
shall see below.

Extrapolating the continuous description of fluid motion to a molecular
scale might be conceptually difficult but unavoidable as far as interfacial
dynamics is concerned. Long-range intermolecular interactions, such as
London--van der Waals forces, still operate on a mesoscopic scale where
continuous theory is justified, but they should be bounded by an inner
cut-off $d$ of atomic dimensions. Thus, distinguishing the first molecular
layer from the bulk fluid becomes necessary even in equilibrium theory.
In dynamic theory, the motion in the first molecular layer can be
described by Eq.~(\ref{slide1}), whereas the bulk fluid obeys hydrodynamic
equations supplemented by the action of intermolecular forces. Equation
(\ref{slide1}) serves then as the boundary condition at the solid surface.
Moreover, at the contact line, where the bulk fluid layer either terminates
altogether or gives way to a monomolecular precursor film, the same slip
condition defines the slip component of the flow pattern, and
Eq.~(\ref{slide1}) can be used to estimate the ``true'' contact angle if it
is assumed that the motion is pure slip at the contact line.

Miller and Ruckenstein \cite{MilRuc74} used the dependence of the
disjoining pressure generated by London--van der Waals forces in a wedge to
compute the ``true'' equilibrium contact angle. This result has been used
by Hocking \cite{Hock93} to set the boundary condition at the contact line
in the hydrodynamic theory, and by Ruckenstein and Dunn \cite{Ruck79} to
compute the slip velocity. Order-of-magnitude estimates show, however, that
at small inclination angles necessary to justify the lubrication
approximation used in hydrodynamic theory the correction to disjoining
pressure due to surface inclination is extremely small, and the ``true''
angle may be formally attained only at distances far below atomic
dimensions. At higher inclination angles, the computation fails
technically, since the interface must be curved, and its form should be
determined by a very complicated integro-differential equation involving
intermolecular interactions as well as viscous stress and surface tension.

We propose to use the kinetic slip condition in the another way, to obtain
a relation between the slip velocity and the thermodynamic potential at the
contact  line by considering the motion at the point where the film thins
down to the minimal thickness $h=d$. This is the advancing edge of a
wetting film, or a retreating edge of a dewetting film, dividing it from
the dry solid surface. If the fluid is not volatile, the motion at this
point should be pure slip, while standard caterpillar motion is retained at
observable macroscopic distances. In the case of an advancing wetting film,
we expect that the leading edge is followed by a thin precursor film where
surface tension is negligible and the action of intermolecular forces
driving the advancing film is balanced by viscous dissipation. The boundary
condition at the leading edge will be then the same Eq.~(\ref{slide1}) with
$u_s$ replaced by the edge propagation speed $U$ and $\nabla w$ computed at
$h=d$ with surface tension neglected.

Another possibility might be to assume that the slip motion at the edge is
driven by the potential drop over the molecular cut-off distance $d$. This
yields, by analogy with Eq.~(\ref{slide1}), the boundary condition at the
contact line
\begin{equation}
  U = - \frac{D}{n kT} \, \frac{w(d)}{ d}, \label{bc}
\end{equation}
where $w(d)$ is the thermodynamic potential of the film of the minimal
thickness; the potential at the dry surface is taken as zero. After $w(d)$
is computed as in the following Section, Eq.~(\ref{bc}) turns into a
condition relating the curvature at the contact line with the propagation
speed. We shall see that this condition leads in fact to non-physical results at small 
propagation velocities. At large velocities, computations using the alternative boundary 
conditions yield practically the same results (see Section
\ref{Sasym}).

\section{Basic equations and scaling} \label{Scale}

We shall use the lubrication approximation, which is formally obtained by
scaling the two-dimensional gradient operator along the solid surface
$\nabla \propto \epsilon$, $\epsilon \ll 1$. Respectively, time is scaled as $\partial_t \propto 
\epsilon^2$, the velocity in the
direction parallel to the solid support as $u \propto \epsilon$ and transverse velocity as $v 
\propto \epsilon^2$. This implies that the
thermodynamic potential $w$ is constant across the layer. The gradient of
$w$ in the direction parallel to the solid support serves as the forcing term in the Stokes 
equation. The velocity profile $u(z)$ across the film verifies
\begin{equation}
\nabla w = \eta u_{zz}, \;\;\; u_z(h)=0, \;\;\; u(d)=u_s,
\label{zstoke}  \end{equation}
where $\eta$ is the dynamic viscosity. We use here the no-stress boundary
condition on the free surface $ z=h$, but replace the usual no-slip
boundary condition on the solid support $u(0)=0$ by the slip condition at
the molecular cut-off distance $d$ with $u_s$ given by Eq.~(\ref{slide1}).
The solution in the bulk layer $ d<z<h $ is
\begin{equation}
u= -\eta^{-1} \left[\lambda^2 +
  h(z- d) -\mbox{$\frac{1}{2}$}(z^2-d^2) \right] \nabla w,
\label{uz}  \end{equation}
where $\lambda =\sqrt{D\eta/n kT}$ is the effective slip length.

The general balance equation for the film thickness $h$, obtained from the
kinematic condition on the free surface, can be presented as a generalized
Cahn--Hilliard equation, where the two-dimensional flux ${\bf j}$ in the
plane aligned with the solid support is proportional to the two-dimensional
gradient of the potential $w$:
\begin{equation}
h_t + \nabla \cdot {\bf j} =0, \;\;\;
{\bf j} = - \eta^{-1} Q(h) \nabla w.
\label{hj}  \end{equation}
The effective mobility $\eta^{-1}Q(h)$ is obtained by integrating
Eq.~(\ref{uz}) across the layer. Including also the constant  slip velocity
$u=-\lambda^2 \eta^{-1}\nabla w$ in the slip layer $0<z<d$, we have
\begin{equation}
Q(h) = \left[ \lambda^2 h + \mbox{$\frac{1}{3}$} (h-d)^3 \right].
\label{qh}  \end{equation}
Since both $\lambda$ and $d$ are measurable on the molecular scale (see the
estimates in the end of this Section), this expression does not differ in a
macroscopically thick layer from the standard shallow water mobility $Q_0=
\frac{1}{3} h^3$, and the correction becomes significant only in the
immediate vicinity of the contact line.

The potential $w$ is computed at the free surface $z=h$. Taking into
account surface tension, gravity, and van der Waals force, it is expressed
as
\begin{equation}
w = - \sigma \epsilon^2 \nabla^2 h +  g\rho(h-\alpha x) -  \frac{A}{6\pi h^{3}},
\label{eqp}  \end{equation}
where $A$ is the Hamaker constant, $g$ is acceleration of gravity, $\rho$
is density, $\sigma$ is surface tension, and $\epsilon\alpha$ is the
inclination angle of the solid surface along the $x$ axis. The dummy small
parameter $\epsilon$ is the ratio of characteristic scales across and along
the layer; the relative scaling of different terms in Eq.~(\ref{eqp}) is
formally consistent when $\sigma = O(\epsilon^{-2})$. Further on, we
suppress the dependence on the second coordinate in the plane, replacing
the Laplacian by $d^2/dx^2$.

In the following, we shall consider the film with a contact line steadily
advancing along the $x$ axis in the negative direction with the speed $U$.
Then Eq.~(\ref{hj}) can be rewritten in the comoving frame, thus replacing
$h_t$ by $U h_x$, and integrated once. Making use of the condition
of zero flux through the contact line to removing the integration constant
yields
 \begin{equation}
  -\eta Uh + Q(h) w'(x) =0.
 \label{hjs} \end{equation}
This equation can be further transformed using $h$ as the independent
variable and $y(h)=h_x^2$ as the dependent variable. We rewrite the
transformed equation introducing the capillary number Ca $ =|U|\eta/\sigma \epsilon^2$,
van der Waals length $a =\epsilon^{-1} (|A|/6\pi\sigma)^{1/2}$, and gravity
length $b=\epsilon(\sigma/g\rho)^{1/2}$:
\begin{equation}
\frac{h\mbox{Ca}}{\sqrt{y}\,Q(h)} + \frac{1}{2}y''(h) - \frac{3a^2}{h^4}
 - \frac{1}{b^2} \, \left(1-\frac{\alpha}{\sqrt{y}}\right) =0.
 \label{main} \end{equation}
%The sign of the first term corresponds to $U<0$.
The boundary condition following from (\ref{slide1}) and balancing
intermolecular forces and viscous dissipation at $h=d$ takes the form
\begin{equation}
 y(d) =\left( d^4\mbox{Ca} / 3\lambda^2a^2\right)^2.
 \label{mbc1} \end{equation}
The alternative boundary condition (\ref{bc}), set at $h=d$, is rewritten,
using Eq.~(\ref{eqp}) and neglecting the gravity term, as
\begin{equation}
  \mbox{$\frac{1}{2}$} y'(d) = \frac{d\mbox{Ca}}{\lambda^2} - \frac{a^2}{d}.
 \label{mbc} \end{equation}

Equation (\ref{main}) contains three microscopic scales $d,a,\lambda$ and a
macroscopic gravity length $b$. The natural choice for $d$ is the nominal
molecular diameter, identified with the cut-off distance in the van der
Waals theory. The standard value \cite{Isr} is $0.165$nm. The slip length
is likely to be of the same order of magnitude. The approximate relation
between viscosity $\eta$ and self-diffusivity $D_m$ in a liquid \cite{fren}
yields $D_m\eta  \approx 10^2 kT/3\pi d$. The surface diffusivity should be
somewhat lower than the diffusivity in the bulk liquid, and with $D/D_m
\approx 0.1$ we have $\lambda \approx d$.

The van der Waals length $a$ depends on the relative strength of
liquid--liquid and liquid--solid interactions. The Hamaker constant for the
pair fluid--solid is defined \cite{Isr} as
\begin{equation}
A = \pi^2 n (C_s n_s - C_f n ) \equiv \pi^2 n^2 \widetilde C,
\label{hamak}  \end{equation}
where $C_s, \ C_f$ are constants in the long-range attraction  potential
$C/r^6$, respectively, for the pairs of fluid--solid and fluid--fluid atoms
removed at the distance $r$; $n_s$ is the solid number density; The effective interaction 
parameter $\widetilde C$ is defined by the above identity. We shall
assume $A>0$, which corresponds to the case of complete wetting. The
estimate for surface tension \cite{Isr} is $\sigma \approx \frac{1}{24} \pi
C_f (n/d)^{-2}$. This gives $a \approx 2 \epsilon^{-1}( \widetilde C
/C_f)^{1/2}$, so that $a = O(d)$ when $\widetilde C /C_f = O(\epsilon^2)$.

\section{Intermediate Asymptotics} \label{Sasym}

In the intermediate region, where $h$ far exceeds the microscopic scales
$d,a,\lambda$ but is still far less than the capillary length $b$, the film
profile is determined by the balance between viscous stress and surface
tension. The asymptotics of the truncated Eq.~(\ref{main}) (with
$d,a,\lambda$, and $b^{-1}$ set to zero) at $h \to \infty$ is 
\begin{eqnarray}
y &\asymp& \left(3\mbox{Ca} \ln \frac{h}{h_0}\right)^{2/3} \cr
&-& 2 \left(\frac{\mbox{Ca}}{9}\right)^{2/3} \ln \ln \frac{h}{h_0}
\left(\ln \frac{h}{h_0}\right)^{-1/3} + \ldots ,
\label{asy}  \end{eqnarray}
where $h_0$ is an indefinite constant. The first term of this asymptotic expression has been 
obtained by Hervet and de Gennes \cite{DGen84}, who have also reported the value of 
$h_0$. This constant can be obtained by
integrating Eq.~(\ref{main}) (with gravity neglected) starting from the boundary condition 
(\ref{mbc1}) or (\ref{mbc}) and adjusting another necessary boundary value
to avoid runaway to $\pm\infty$. There is a unique heteroclinic trajectory approaching the 
asymptotics (\ref{asy}). It is very sensitive to the
initial conditions as well as to the molecular-scale factors operating
close to the contact line. The growth of the inclination angle is never saturated, as long as 
macroscopic factors (gravity or volume constraint) are not taken into account.

\begin{figure}
\begin{tabular}{c}
(a)\\
\psfig{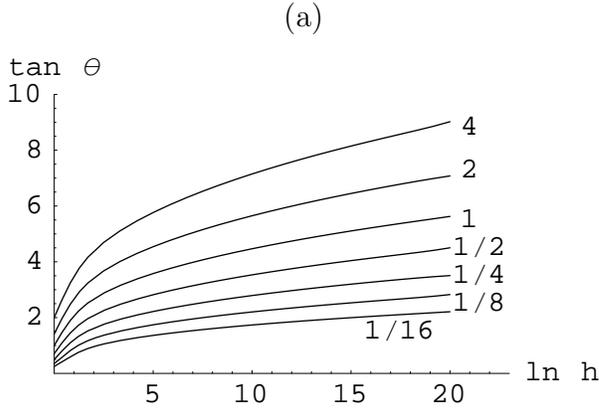} \\
(b) \\
\psfig{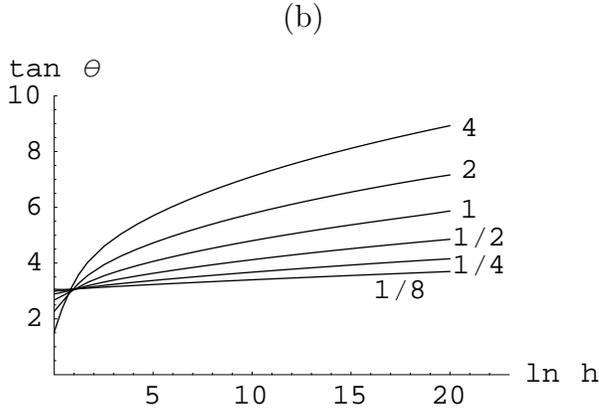}
\end{tabular}
\caption{Dependence of the local surface inclination $\tan \theta$ on the
local film thickness at different values of the capillary number Ca computed using the 
boundary condition (\ref{mbc1}) (a) and (\ref{mbc}) (b). The numbers at the curves show 
the values of Ca. Other
parameters used in all computations are $\lambda=1$, $a=1/\sqrt{3}$,}
\label{f1}
\end{figure}

Equation (\ref{main}) can be integrated using the shooting method: either starting from the 
boundary condition (\ref{mbc1}) and adjusting $y'(d)$ or
starting from the boundary condition (\ref{mbc}) and adjusting $y(d)$ to arrive at the 
required asymptotics at $h \to \infty$. Further on, we will measure all
lengths in the molecular units and set $d$ to unity. The solution in the
intermediate region depends on the physical parameters $a,\lambda$ as well
as on the capillary number Ca  that includes the propagation speed $U$. The
latter's impact is most interesting for our purpose.  Examples of the
computed dependence of the inclination angle  $\theta=\sqrt{y(h)}$ on the
local film thickness $h$ using the boundary condition (\ref{mbc1}) at
different values of Ca are given in Fig.~\ref{f1}.

The curves using the boundary condition
(\ref{mbc}) in Fig.~\ref{f1}b, show peculiar (apparently, non-physical) reversal of the 
dependence of the inclination angle on Ca, resulting in an {\it increase\/} of the ``true" 
contact angle $\theta(d)$ with decreasing velocity. Indeed, at $U \to 0$ this condition yields 
a spurious balance between intermolecular
forces and surface tension leading to an {\it unstable\/} stationary state,
similar to the erroneous inference of a wetting film with the right
contact angle in Ref.~\cite{DGen90} discussed in our earlier paper
\cite{prb}. The anomaly, however, quickly disappears at observable distances.

\begin{figure}
\psfig{figure=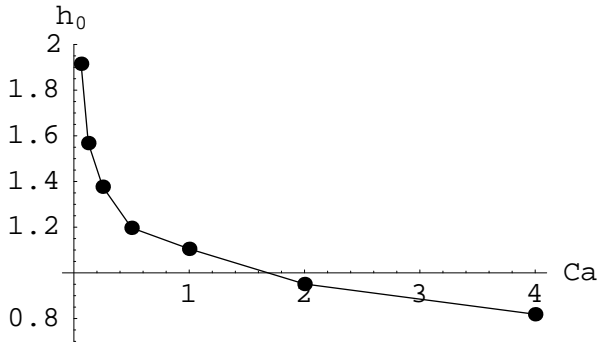,width=8.0cm}
\caption{Dependence of $h_0$ on Ca computed using the data from
Fig.~\protect{\ref{f1}}a. }
\label{f2}
\end{figure}

The curve segments at $h\gg 1$ can be fit to the asymptotic formula (\ref{asy}) to obtain 
the integration constant $h_0$. It should be noted that the asymptotic formula (\ref{asy}) 
can be used only when
$h$ is {\it logarithmically} large, and the convergence, as estimated by the second term, is 
slow; therefore $h_0$ can be only obtained approximately from the computed profiles. The 
dependence of $h_0$ on Ca based on Fig.~\ref{f1}a is shown in Fig.~\ref{f2}. We see 
here a rather strong variation of the integration constant, unlike a single ``universal'' value 
reported in Ref.~\cite{DGen84}.

\section{Draw-down of a meniscus} \label{Smen}

The simplest stationary arrangement including gravity is realized when an
inclined plane, dry at $x \to -\infty$ slides in the direction of a wetting
layer. Solving Eq.~(\ref{main}) with the same boundary
condition (\ref{mbc}) as before brings now to the asymptotics
$y=\sqrt{\alpha}$ at $h \to \infty$ that corresponds to a horizontal layer.

\begin{figure}
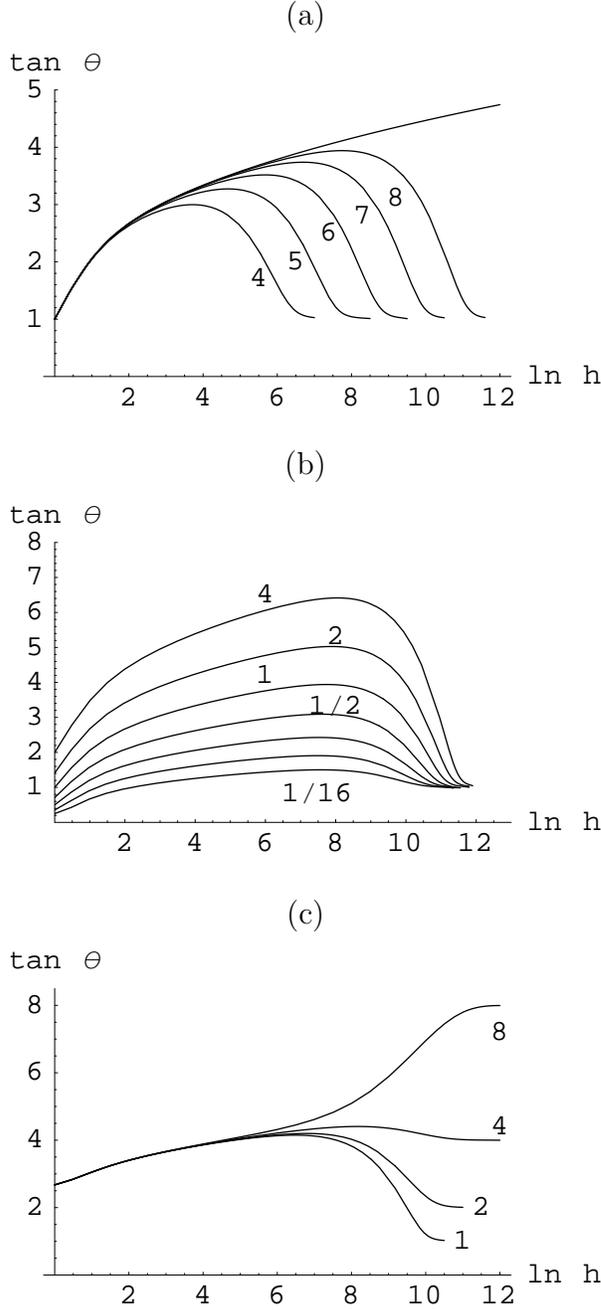

\begin{tabular}{c}
(a)\\
\psfig{figure=Slip/fig3a.eps,width=8.0cm} \\
(b) \\
\psfig{figure=Slip/fig3b.eps,width=8.0cm} \\
(c) \\
\psfig{figure=Slip/fig3c.eps,width=8.0cm}
\end{tabular}
\caption{Dependence of the local surface inclination angle $\theta$ on the
film thickness  (a) at Ca$=1$, $\alpha=1$ and different values of the
gravity length $b$; (b) at $b=10^4$ and different values of the capillary
number Ca; (c) at Ca$=1$, $b=10^4$ and different values of the asymptotic
inclination angle $\alpha$. The numbers at the curves show the values, respectively, of 
$2\log b$, Ca and $\alpha$. Other parameters used in all computations are
$\lambda=1$, $a=1/\sqrt{3}$.}
\label{f3}
\end{figure}

The curves $y(h)$ seen in Fig.~\ref{f3}a and Fig.~\ref{f3}c all depart from
the intermediate
asymptotic curve obtained for infinite $b$ as in the preceding Section.
However, due to extreme sensitivity of the shooting method to the choice of
the missing initial value, one has to integrate from the outset the
full equation rather than trying to start integration from some point on
the intermediate asymptotic curve. One can see that the maximum inclination
angle (which may be identified with the ``visible" contact angle) grows as
$b$ decreases. This increase is, however, not pronounced when the initial
incline (identified with the ``true" contact angle) is high. One can
distinguish therefore between two possibilities: first, when the main
dissipation is due to kinetic resistance in the first monomolecular layer
that raises $y(d)$, and second when the viscous dissipation prevails and
the inclination angle keeps growing in the region of bulk flow. Take note
that even in the latter case the region where the inclination and curvature
are high are close to the contact line when measured on a macroscopic
scale.

\begin{figure}
\psfig{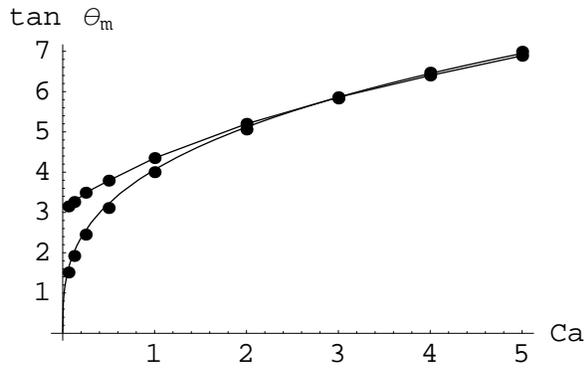}
\caption{Dependence of the visible contact angle $\theta_m$ on Ca.}
\label{fG}
\end{figure}

Figure \ref{fG} shows the dependence of the ``visible" contact angle
$\theta_m$, defined as the maximum inclination angle and observed in the range where the 
gravity-dependent curves depart from the intermediate asymptotics, on the capillary
number Ca. The lower curve is a fit $\theta \propto \mbox{Ca}^{1/3}$ to the data of 
Fig.~\ref{f3}b. The points of the upper
curve are computed in a similar way using the boundary condition
(\ref{mbc}). The first result appears to be more physically reasonable,
since the angle drops close to zero at small flow velocities, while in the
alternative computation it remains finite (see also the discussion in the preceding Section). 
The proportionality of the inclination angle to Ca$^{1/3}$ (which leads to the well-known 
Tanner¹s law of spreading \cite{DGen85}) is a property of the intermediate asymptotics 
(\ref{asy}) that  can be deduced from scaling \cite{prb}, although the universality is 
slightly impaired by the dependence of the integration constant $h_0$ on velocity seen in 
Fig.~\ref{f2}.. The one-third law is inherited by the dependence $\theta_m$(Ca), since the 
inclination angle reaches its maximum while the gravity-dependent profile is still close to 
the intermediate asymptotic curve.

\begin{figure}
\psfig{figure=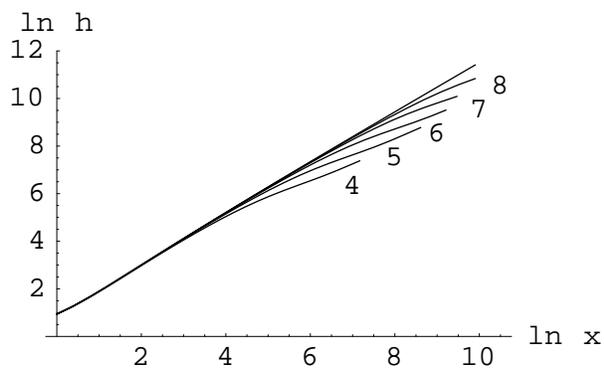,width=8.0cm}
\caption{The shape of the meniscus for different values of $b$. 
The numbers at the curves 
show the values of $2\log b$.}
\label{f4}
\end{figure}

\begin{figure}
\psfig{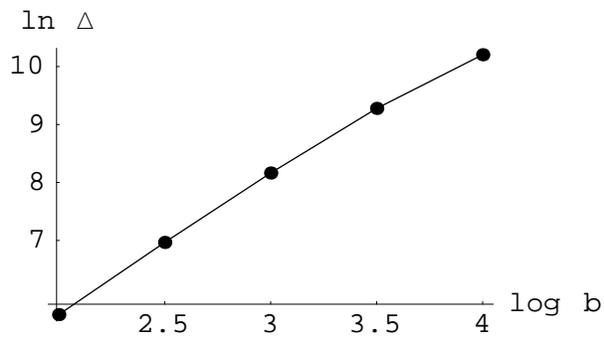}
\caption{The dependence of the draw-down
length $\Delta$ on $\log b$.}
\label{f5}
\end{figure}

Fig.~\ref{f4} shows the actual shape of the meniscus obtained by integrating
the equation $h'(x)=\sqrt{y(h)}$, $h(0)=d$. The dependence of the draw-down
length $\Delta$ (computed as the difference between the actual position of
the contact line and the point where the continuation of the asymptotic
planar interface hits the solid surface) on the gravity length is shown in
Fig.~\ref{f5}.

\section{Conclusion}

It comes, of course, as no surprise that introducing a molecular cut-off and applying a 
kinetic slip condition to the first molecular layer resolves the notorious singularities of 
hydrodynamic description. The hydrodynamic singularities are eliminated, however, only 
at molecular distances, and are still felt in sharp interface curvatures at microscopic 
distances identified here as the intermediate asymptotic region. The computations are eased 
considerably when non-physical divergence of both viscous stress and attractive Lennard--
Jones potential beyond the cut-off limit are eliminated. As a result, the stationary equations 
can be solved by shooting method with reasonable accuracy in a very wide range extending 
from molecular to macroscopic scales, and the ``true" contact angle at the cut-off distance 
can be defined unequivocally. 

The ``true" angle (unobservable by available techniques) depends on the slip length as well 
as on the edge propagation speed, but not on gravity or asymptotic inclination angle. These 
macroscopic factors influence, however, the ``visible" contact angle observed in the 
interval where the actual film profile departs from the intermediate asymptotic curve. Since 
the latter¹s location, though not shape, depends on the molecular-scale factors, as well as 
on the cut-off distance, the visible angle depends on both molecular and macroscopic 
factors. Thus, the lack of simple 
recipes for predicting the value of dynamic contact angle is deeply rooted in the mesoscopic 
character of the contact line.

\acknowledgments
This research has been supported by the Israel Science Foundation. LMP
acknowledges partial support from the Minerva Center for Nonlinear Physics
of Complex Systems.

%%%%%%%%%%%%%%%%%%%%%%%%%%%%%%%%%%%%%%%%%%%
%%%%%%%%%%%%%%%%%%%%%

%\newpage

%\newpage

%\newpage

%\newpage

%\newpage

\end{document}